

\input phyzzx

\overfullrule=0pt

\let\refmark=\NPrefmark 
\def\define#1#2\par{\def#1{\Ref#1{#2}\edef#1{\noexpand\refmark{#1}}}}
\def\con#1#2\noc{\let\?=\Ref\let\<=\refmark\let\Ref=\REFS
         \let\refmark=\undefined#1\let\Ref=\REFSCON#2
         \let\Ref=\?\let\refmark=\<\refsend}

\define\RCHOUSCH
S. Chaudhuri and J. A. Schwarz, Phys. Lett. {\bf B219} (1989) 291.

\define\RHORAVA
P. Horava, Phys. Lett. {\bf B278} (1992) 101.

\define\RVENBH
M. Gasperini, J. Maharana and G. Veneziano, preprint CERN-TH.6634/92
(hepth/9209052).

\define\RHS
C. Hull and B. Spence, Phys. Lett. {\bf B232} (1989) 204.

\define\RROCVER
M. Rocek and E. Verlinde, Nucl. Phys. {\bf B373} (1992) 630.

\define\RJJM
I. Jack, D. Jones, N. Mohammedi and H. Osborn, Nucl. Phys. {\bf B332}
(1990) 359.

\define\RGAZU
M. Gaillard and B. Zumino, Nucl. Phys. {\bf B193} (1981) 221.

\define\RODDNEW
J. Horne, G. Horowitz and A. Steif, Phys. Rev. Lett. {\bf 68}
(1992) 568;
J. Panvel, Phys. Lett. {\bf B284} (1992) 50.

\define\RWIT
E. Witten, Phys. Rev. {\bf D44} (1991) 314;
C. Nappi and E. Witten, preprint IASSNS-HEP-92/38 (hepth/9206078)
and references therein.

\define\RKUM
S. Kar, S. Khastagir and A. Kumar, Mod. Phys. Lett. {\bf A7} (1992) 1545;
S. Kar and A. Kumar, preprint IP-BBSR-92-18 (hepth/9204011);
J. Maharana, preprint CALT-68-1781 (hepth/9205016);
S. Khastagir and J. Maharana, preprint IP-BBSR-92-38 (hepth/9206017).

\define\RALOK
A. Kumar, preprint CERN-TH.6530/92 (hepth/9206051).

\define\RMAHSCH
J. Maharana and J. Schwarz, preprint CALT-68-1790 (hepth/9207016).

\define\RODD
S. Ferrara, J. Scherk and B. Zumino, Nucl. Phys. {\bf B121} (1977) 393;
E. Cremmer, J. Scherk and S. Ferrara, Phys. Lett. {\bf B68} (1977) 234;
{\bf B74} (1978) 61;
E. Cremmer and J. Scherk, Nucl. Phys. {\bf B127} (1977) 259;
E. Cremmer and B. Julia, Nucl. Phys.{\bf B159} (1979) 141;
M. De Roo, Nucl. Phys. {\bf B255} (1985) 515; Phys. Lett. {\bf B156}
(1985) 331;
E. Bergshoef, I.G. Koh and E. Sezgin, Phys. Lett. {\bf B155} (1985) 71;
M. De Roo and P. Wagemans, Nucl. Phys. {\bf B262} (1985) 646;
L. Castellani, A. Ceresole, S. Ferrara, R. D'Auria, P. Fre and E. Maina,
Nucl. Phys. {\bf B268} (1986) 317; Phys. Lett. {\bf B161} (1985) 91;
S. Cecotti, S. Ferrara and L. Girardello, Nucl. Phys. {\bf B308} (1988)
436;
M. Duff, Nucl. Phys. {\bf B335} (1990) 610.

\define\RVENEZIANO
G. Veneziano, Phys. Lett. {\bf B265} (1991) 287;
K. Meissner and G. Veneziano, Phys. Lett. {\bf B267} (1991) 33; Mod. Phys.
Lett. {\bf A6} (1991) 3397;
M. Gasperini, J. Maharana and G. Veneziano, Phys. Lett. {\bf B272} (1991)
277;
M. Gasperini and G. Veneziano, Phys. Lett. {\bf B277} (1992) 256.

\define\RGIVROC
A. Giveon and M. Rocek, Nucl. Phys. {\bf B380} (1992) 128;

\define\RGIVEON
A. Giveon and A. Pasquinucci, preprint IASSNS-HEP-92/35
(hepth/9208076).

\define\RSEN
A. Sen, Phys. Lett. {\bf B271} (1991) 295; {\bf 274} (1991) 34.

\define\RFAWAD
S.F. Hassan and A. Sen, Nucl. Phys. {\bf B375} (1992) 103.

\define\RROTSTR
A. Sen, Phys. Rev. Lett. {\bf 69} (1992) 1006; preprint TIFR-TH-92-39
(hepth/9206016) (to appear in Nucl. Phys. B).

{}~\hfill\vbox{\hbox{TIFR-TH-92-61}\hbox{hep-th/9210121}\hbox{October,
1992}}\break

\title{MARGINAL DEFORMATIONS OF WZNW AND COSET
MODELS FROM $O(d,d)$ TRANSFORMATION}

\author{S.F. Hassan and Ashoke Sen\foot{e-mail addresses:
FAWAD@TIFRVAX.BITNET, SEN@TIFRVAX.BITNET}}

\address{Tata Institute of Fundamental Research, Homi Bhabha Road, Bombay
400005, India}

\abstract

We show that $O(2,2)$ transformation of $SU(2)$ WZNW model gives rise to
marginal deformation of this model by the operator $\int d^2 z J(z)\bar
J(\bar z)$ where $J$, $\bar J$ are $U(1)$ currents in the Cartan
subalgebra.
Generalization of this result to other WZNW theories is discussed.
We also consider $O(3,3)$ transformation of the product of
an $SU(2)$ WZNW model and a gauged $SU(2)$ WZNW model.
The three parameter set of models obtained after the transformation is
shown to be the result of first deforming the product of two $SU(2)$ WZNW
theories by marginal operators of the form $\sum_{i,j=1}^2 C_{ij} J_i
\bar J_j$, and then gauging an appropriate $U(1)$ subgroup of the theory.
Our analysis leads to a general conjecture that $O(d,d)$ transformations of
any WZNW
model correspond to marginal deformation of the WZNW
theory by an appropriate combination ofleft and right moving currents
belonging to the Cartan subalgebra; and $O(d,d)$ transformations of a
gauged WZNW model can be identified to the gauged version of such
marginally deformed WZNW models.

\chapter {Introduction}

$O(d,d)$ transformations\RGAZU\RODD\ have been used in many recent
papers to
generate new classical solutions of string theory equations of motion from
known
ones\con\RVENEZIANO\RSEN\RFAWAD\RKUM\RODDNEW
\RROTSTR\RALOK\RGIVEON\RMAHSCH\RVENBH\noc.
In this paper we shall discuss a new application of these transformations,
namely, generating marginal deformations of Wess-Zumino-Novikov-Witten
(WZNW) and coset models.

The motivation for studying marginally deformed WZNW models and their
gauging
is as follows.  The propagation of a string in a background is described
by a conformally invariant non-linear sigma model in two dimensions.  The
background is constructed as a solution to the vanishing beta function
equations which enforce the condition of conformal invariance
perturbatively . The low energy equations of motion for the background
fields have been solved in various situations and a wide range of
solutions has been obtained. Particular examples are solvable conformal
field theories like WZNW models, or the
"blackhole" type
solutions obtained by gauging WZNW models\RWIT.
In general, these models form a very small subset in the space of all
conformal field theories, and it is of particular interest to know the
interrelation between various exactly solvable conformal field theories of
this kind, e.g. the question of whether they can be obtained from each
other by marginal deformations.
In order to address such questions, we need to know the form of the
$\sigma$-model action of the theory after a finite marginal deformation,
and this is precisely the question we address in this paper.

It has been argued\RCHOUSCH\ that  WZNW
theories have exact marginal
deformations,
generated by operators of the form $\sum_{i,j} C_{ij}\int d^2z J_i(z) \bar
J_j(\bar z)$ where $J_i$ and $\bar J_i$ are holomorphic and
anti-holomorphic currents belonging to the Cartan subalgebra.
The $\sigma$-model action of the corresponding perturbed theory can be
written down easily to first order in the perturbing parameter, but there
is no general method for writing it down for a finite value of the
perturbation parameter.
We shall show that $O(d,d)$ transformations provide a way out of this
problem.
More specifically, we shall show that if we start from an unperturbed
SU(2) WZNW theory, and perform a (finite) $O(2,2)$ transformation on it,
the result is a $\sigma$-model describing an SU(2) WZNW model deformed by
the marginal operator $\int d^2 z J(z) \bar J(\bar z)$.
The form of the $\sigma$-model is exact to all orders in the perturbing
parameter, but only to lowest order in $1/k$, where $k$ is the central
charge of the current algebra.
(This shortcoming is only due to the fact that the explicit $O(d,d)$
transformation rules are known only to lowest order in the derivatives.)
We also consider the generalization of this result for more general WZNW
theories; in particular we discuss the case of $SU(2)\otimes SU(2)$
theory, and show that the four parameter family of marginal
deformations in this theory is again given by $O(4,4)$ transformation of
the unperturbed theory.

Besides the WZNW models, another type of conformal field theories have been
the subject of intense investigation in recent years; these are the
coset models and are obtained by gauging one or more subgroups of some WZNW
model.
In this context, a natural question to ask would be, what kind of models
can we get if we gauge a marginally deformed WZNW model.
We show that the answer is reasonably simple; these models can be
identified with the $O(d,d)$ deformations of the gauged unperturbed WZNW
models.
In other words, we show, through various examples, that gauging an $O(d,d)$
transformed WZNW model generates an $O(d,d)$ transformed coset model.

The paper is organized as follows.
In section 2 we consider an $O(2,2)$ transformation of an $SU(2)$ WZNW
model,
and show that the resulting $\sigma$-model can be identified to the
marginal deformation of the original model by the $\int d^2 z J(z)\bar
J(\bar z)$ perturbation.
We also consider $O(4,4)$ transformation of $SU(2)\otimes SU(2)$ WZNW
model and
show that the result can be identified to marginal deformation of the
original theory by perturbations of the form $\int d^2 z J_i(z)\bar
J_j(\bar z)$ ($1\le i,j\le 2$).
We then discuss $O(d,d)$ transformations of more general WZNW models, and
their
relation to marginal deformations of these models.

In sect.3 we consider $O(3,3)$ transformations of the product of an $SU(2)$
WZNW model, and an $SU(2)/U(1)$ coset model.
The result is a three parameter family of conformal field theories.
The physical interpretation of this conformal field theory is provided by
the analysis of sect.4, where we start with an $O(4,4)$ transformed
$SU(2)\otimes SU(2)$ WZNW
model (which, by the result of
sect.2 is a marginal deformation of the $SU(2)\otimes SU(2)$ WZNW model),
and then gauge a particular $U(1)$ subgroup of the model.
The result is shown to be the model of sect.3, thereby showing that an
$O(d,d)$ transformation of the coset model is equivalent to gauging a
marginally deformed WZNW model.
We also comment on the observations of refs.\RALOK\ and \RVENBH\ in the
context of $SU(2)$ WZNW model.

We conclude in sect.5 with a summary of the results and some comments.

\def\odd{$O(d,d)$}
\def\ott{$O(2,2)$}
\def\thl{\theta_L}
\def\thr{\theta_R}
\def\th{\theta}
\def\thwt{\widetilde\theta}
\def\tht{\tilde\theta}
\def\twt{\widetilde t}
\def\tt{\tilde t}
\def\cp{\cos^2 {\phi\over 2}}
\def\ct{\cos^2 {r\over 2}}
\def\sp{\sin^2 {\phi\over 2}}
\def\st{\sin^2 {r\over 2}}
\def\cosal{\cos \alpha}
\def\sal{\sin \alpha}
\def\cosbe{\cos \beta}

\def\cg{\cos \gamma}
\def\sg{\sin \gamma}
\def\cgsq{\cos ^2 \gamma}
\def\sgsq{\sin ^2 \gamma}
\def\Delinv{{1\over \Delta}}
\def\d{D^{-1}}
\def\del{\partial}
\def\delb{\bar \partial}
\def\Jbar{{\bar J}}
\def\wij{\del_{[i}w_{j]}}
\def\e{\epsilon}
\def\half{{1\over 2}}
\def\g{{\scriptstyle\rm gauged}}
\def\p{\partial}
%
%

\chapter{\odd\ Trasformations and Marginal Perturbations in the
WZNW Models}

To explore the connection between \odd\ transformations and marginal
perturbations in WZNW models, we first consider an $SU(2)$-WZNW model
defined by the action
$$
S[g]={ {k\over 8 \pi}\int_{\del B}d^2 x\, Tr(\del_\mu g^{-1} \del^\mu g)
+{k\over 12 \pi}\int_B d^3 x\, \e^{ijk}\, Tr(\del_i g g^{-1} \del_j g
g^{-1} \del_k g g^{-1})}\eqn\wzw
$$
where $g$ is an $SU(2)$ group element and the constant $k$ specifies
the level of the associated Kac-Moody (KM) algebra. $B$ is a solid ball in
three
dimensions with boundary $\del B$. This model describes a conformal field
theory (CFT) of
central charge $c={3k\over{k+2}}$ and gives the well known {\it2-d}
blackhole on gauging.  If we parametrize $g$ as
$$
g=e^{i\thl \sigma_2 /2} e^{i \phi \sigma_1 /2} e^{i \thr
\sigma_2 /2}\eqn\param
$$
then the action takes the form
$$
S[\phi,\thl,\thr]={k\over 2 \pi}\int d^2z(\delb\phi\del\phi+\delb \thl
\del \thl+\delb \thr \del \thr+2\cos\phi \delb\thl \del \thr)\eqn\suto
$$
which is related to the level $k$ $SL(2,R)$ model by the replacements
$\phi \rightarrow ir$, $k \rightarrow  -k$.
The action \suto\ has chiral invariances
$$
\eqalign{
{\delta \thl}&= v_L(z);\hphantom{0} \qquad \delta \thr=0 \cr
{\bar \delta} \thl&=0;\hphantom{v_L(z)} \qquad \bar \delta \thr={\bar
v}_R (\bar z)\cr}
\eqn\chirinv
$$
which give rise to the conserved chiral currents
$$
\eqalign{
J&={1\over 2}k(\del \thl+\cos \phi\,\del \thr)\cr
\Jbar&={1\over 2}k(\delb \thr+\cos \phi \,\delb \thl)\cr}
\eqn\curr
$$
respectively.

A small deformation of the $SU(2)$-WZNW model can be obtained by adding a
marginal perturbation of the form
$$
O={\delta\lambda \over 2\pi} \int d^2z\,J \Jbar \eqn\pert$$
to the action  \suto. The integrability of this perturbation
\foot{In a CFT with a KM symmetry, all marginal
operators of the form $\sum_{i,j} C_{ij} J_i \Jbar_j$, where $J_i$, $\bar
J_i$ belong to the Cartan subalgebra, are integrable\RCHOUSCH.}
indicates the
existence of a continuous family of CFT's of the same central charge,
parametrized by $
\lambda$, such that
two theories corresponding to two adjacent values of $\lambda$
are related
by a generalization of the operator \pert. Let us denote the action of
the deformed theory by $S_{(\lambda)}$. Except for some relatively simple
cases,  like the theory of a free boson compactified on a circle, the form
of $S_{(\lambda)}$ cannot be obtained  in a straightforward way. In the
following we show that for WZNW models a solution to this problem is
provided by \odd\ transformations.

The WZNW models are special cases of non-linear $\sigma$-models
which are conformally invariant.  As a consequence,  the
$\sigma$-model coupling constants in these theories automatically satisfy
the zero $\beta$-function conditions and correspond to specific
classical configurations of background fields described by
the string theory  effective action.
It has  been shown\RVENEZIANO-\RMAHSCH\ that the string theory low-energy
effective
action, when restricted to background fields that are
independent of
$d$  of the space dimensions, is invariant under an \odd\ group
of transformations. Since this group of transformations is not necessarily
a symmetry of the unrestricted action, it can relate classical
field configurations  which are not equivalent,  giving rise to
different conformally invariant world-sheet theories. In the
following
we briefly describe the action of this group on the background
fields $G_{\mu\nu}, B_{\mu\nu}$ and $\Phi$ of the closed bosonic
string theory. For all our purposes in the present paper it is
sufficient to consider backgrounds of the form
\foot{An extention to general backgrounds is given in ref.\RFAWAD\ in
the context of heterotic string theory}
$$
{G=\pmatrix{\widetilde G_{\alpha\beta}&0\cr
                       0&\widehat G_{mn}\cr}} \quad, \quad
 {B=\pmatrix{\widetilde B_{\alpha\beta}&0\cr
                       0&\widehat B_{mn}\cr}} \eqn\bac
$$
where the indices $m$ and $n$ span the $d$-dimensions on which the fields
do not depend. Let us now construct a $2d\times 2d$ matrix $M$ as
$$
M=\pmatrix{\widehat G^{-1}&-\widehat G^{-1}\widehat B\cr
          \widehat B\widehat G^{-1}&\widehat G-\widehat
B\widehat G^{-1}\widehat B\cr}\eqn\M
$$
In terms of this matrix and the dilaton field $\Phi$, the action of
the \odd\ group is given by
$$
{\eqalign{
M  &\longrightarrow M'=\Omega M \Omega^T \cr
\Phi &\longrightarrow \Phi'= \Phi+{1\over
2}\ln\left[{\det{\widehat G'}\over {\det \widehat
G}}\right]}}\eqn\mphi
$$
where $\Omega$ is an \odd\ group element defined by
$$
\Omega \, \pmatrix{0&1_d \cr 1_d&0\cr} \,\Omega^T=\pmatrix{0&1_d \cr
1_d&0 \cr}
\eqn\efawadone
$$
The components of the fields $G$ and $B$ which do not appear in
$M$ remain unchanged under the transformation.

An $\Omega$ lying in the $O(d)\otimes O(d)$ subgroup of \odd\
can be parametrized as
$$
\Omega_1={1 \over 2} \pmatrix{R+S&R-S \cr R-S&R+S \cr}
\eqn\rspara
$$
with $R,  S \in O(d)$. The diagonal part $R=S$ corresponds to
rotations in $d$ -dimensions.
Thus the non-trivial transformations are generated by matrices of the form
$\Omega_1$ modulo the diagonal subgroup; they form a coset $O(d)\otimes
O(d)/O(d)$.
A general $O(d,d)$ transformation may be expressed as the product of an
element of the coset $O(d)\otimes O(d)/O(d)$ (which, in turn, may be
labelled by an element of the group $O(d)$ by making the specific choice
$S=R^T$), and the group
generated by matrices of the form\RSEN:
$$
\Omega_2 = \pmatrix{(A^T)^{-1}&0 \cr 0&A} \, , \qquad
\Omega_3 = \pmatrix{1_d&0 \cr C&1_d \cr}
\eqn\acpara
$$
where $A$ and $C$ are constant matrices in $d$-dimensions and
$C$ is antisymmetric. They generate transformations of the form
$ \widehat G \rightarrow A \widehat G A^T ,
  \widehat B \rightarrow A \widehat B A^T $ and $ \widehat B
\rightarrow \widehat B + C$.
These
are implemented by general coordinate transformations of the
form $ X'^m = {A^m}_n X^n$ and gauge transformations  of
$B_{mn}$ with gauge parameter $\Lambda_m = C_{mn} X^n$. The
elements of the form \rspara\ with $R=S^T$,  on
the other hand,  act non-linearly on the background fields and,
for non-compact coordinates,  are the only elements which give
rise to inequivalent backgrounds.
For WZNW models the coordinates are compact, and a new
background obtained by a transformation of coordinates is not equivalent
to the old background if we assume the same periodicity for the new and
the old coordinates.
In fact, since these coordinates are angular variables, the new background
generically will correspond to a singular metric even if the original
metric was non-singular.\foot{A trivial example of this is the metric
$dr^2 +r^2 d\theta^2$ in polar coordinates. If we define $\theta'
=\theta/C$, and consider the case where $\theta'$ is an angular variable
with period $2\pi$, then the new metric $dr^2 + C^2 d(\theta')^2$ has
conical singularity at  $r=0$.}
As we shall see, the field configuration obtained by transforming the WZNW
theory by the elements of the form given in eq.\rspara\ with $R=S^T$
is quite often
singular, and needs to be followed by a general coordinate transformation
in order to give a non-singular metric.

Although \odd\ symmetry
survives to all orders in the $\sigma$-model perturbation
theory\RSEN\RFAWAD,  the
explicit form given above is correct only to the lowest order in the
$\sigma$-model loop expansion parameter.
In applications to WZNW models,  therefore,  the
explicit results obtained are valid in the large-$k$ limit. We
will always assume this to be the case.

Returning to the action \suto\ we note that the target space metric and
antisymmetric tensor fields of the $\sigma$-model
are independent of the two coordinates $\thl$ and $\thr$.
This allows us to perform an \ott\ transformation on the model.
We first define new coordinates
$$
\th = {1 \over 2}(\thl-\thr) \,,  \qquad
\thwt = {1 \over 2}(\thl+\thr)
\eqn\efawadtwo
$$
in terms of which the metric $G_{\mu \nu}$ is diagonal. After
shifting the $B$-field by a constant matrix,  the action \suto\
becomes
$$
S(\phi,\th,\thwt)={k \over 2 \pi}\int d^2 z[\,{1 \over 4}
\delb \phi \del \phi +\sp \delb \th \del \th + \cp \delb
\thwt \del \thwt + \cp (\delb \th \del \thwt - \delb \thwt
\del \th)\,]\eqn\szero
$$
from which,  we can read off the $\sigma$-model coupling constants as
$$
G=\pmatrix{{k/4}&0&0\cr0&k\sp&0\cr 0&0&k\cp\cr} \,,\quad
B=\pmatrix{0&0&0\cr 0&0&k\cp \cr 0&-k\cp&0 \cr}
\eqn\efawadthr
$$
and $ \Phi = 0$.
Since $\th$ and $\thwt$ have periodicities of $2\pi$,  the metric
does not have coordinate singularities at $\phi = 0$ and
$\phi=\pi$.

Now we consider a transformation of the above
backgrounds by the elements $\Omega_1$ of the \ott\ group as
given in \rspara,   with the parametrization
$$
R = S^{-1} = \pmatrix{\cosal &\sal \cr -\sal &\cosal\cr}
\eqn\efawadfour
$$
The new metric has conical singularities at $\phi = 0$ and $\pi$
which are removed by scalings $\th \rightarrow \th/(
\cosal-k\sal)$ and $\thwt \rightarrow \thwt/\cosal$ (This
restores the periodicity of $\th$ and $\thwt$ to $2\pi$). We
also make a further transformation $B_{\tht\theta} \rightarrow
B_{\tht\theta} + L $,  where $L$ is a constant given by
$$ L = \cosal (k\cosal + \sal) \eqn\l $$
This corresponds to locally adding a total derivative term to the action
and does not change the $\beta$-function of the theory.
We will elaborate more on this choice of $L$ while discussing
the gauging of the \ott\ transformed model.
The new $\sigma$-model action obtained after these
transformations is
$$
{\eqalign{
S_{(\alpha)}(\phi,\th,\thwt)&={k \over 2 \pi}\int d^2 z[\,{1
\over 4} \delb \phi \del \phi +\Delinv (\cosal-k\sal)^2 \sp \delb
\th \del \th
\cr  & \qquad\qquad
+ \Delinv \cos ^2 \alpha \cp \delb \thwt \del \thwt
 -\Delinv \cos ^2\alpha \sp (\delb \th \del \thwt - \delb \thwt
\del \th)\,]}}\eqn\salpha
$$
with a dilaton field $\Phi$ given by\foot{There is a freedom of adding an
overall constant to the expression for $\Phi$, which we have not displayed
explicitly.
While comparing different solutions we should keep this in mind.}
$$
\Phi =-\ln \Delta \eqn\dilalpha
$$
where,
$$
\Delta = \cos^2 \alpha +k\,(k\sin^2 \alpha -2\sal \cosal)
\cp
\eqn\efawadfive
$$
Although the explicit form of the transformed action is valid
only in the large $k$ limit, the existence of the \ott\
transformation, and hence of the transformed action can be shown
to all orders in $k$.
The action \salpha\ along with the dilaton \dilalpha\ describes a
one-parameter ($\alpha$) family of CFT's which include the $SU(2)$ WZNW
model ($\alpha = 0$). For small $\alpha$ the change in the action
\szero\ under the \ott\ transformation is
$$
\delta S = S_{(\alpha)}-S_{(0)} ={\alpha \over \pi} \int d^2z\,J
\Jbar
\eqn\efawadsix
$$
where,  $J$ and $\Jbar$ are the currents \curr\ now written in
terms of $\th$ and $\thwt$. A comparison with \pert,  for
$\lambda = 2 \alpha$,  shows that the \ott\ transformed action
for small $\alpha$ is the same as the action obtained after a
marginal perturbation. To generalize this result to finite
$\alpha$,  we notice that
$$
S_{(\alpha + \delta \alpha)}=S_{(\alpha)} + {\delta \alpha \over
\pi} \int d^2z\,[{1\over \Delta ^2} \cosal(\cosal-k\sal)  J \Jbar]
\eqn\deltas
$$
with $J$, $\bar J$ as defined in eq.\curr.
The equations of motion for $\th$ and $\thwt$,   obtained from
$S_{(\alpha)}$,  are $\delb (J/\Delta) = \del (\Jbar /\Delta)=0$.
The significance of these equations may be understood in the following
way. The action $S_{(\alpha)}(\phi,\th,\thwt)$ has chiral invariances
$$
\eqalign{
{\delta \th}&=\half {1\over (\cosal-k\sal)^2} v(z),\hphantom{-}
\qquad \delta \thwt=\half {1\over \cos^2 \alpha} v(z) \cr
\bar \delta \th&=-\half {1\over(\cosal-k\sal)^2}\bar v(\bar z),
\qquad \bar \delta \thwt=\half {1\over \cos^2 \alpha}\bar v (\bar z)\cr}
\eqn\chalinv
$$
The corresponding conserved chiral currents have the form
$$
\eqalign{
J_{(\alpha)}&={k\over \Delta}(\sp\del\th+\cp\del\thwt)=J/(\Delta) \cr
\Jbar_{(\alpha)}&={k\over \Delta}(-\sp\delb\th+\cp\delb\thwt)=
\Jbar/(\Delta)}
\eqn\curral
$$
In terms of $J_{(\alpha)}$ and $\Jbar_{(\alpha)}$ \deltas\ becomes
$$
S_{(\alpha + \delta \alpha)}=S_{(\alpha)} + {\delta \alpha \over
\pi}\cosal(\cosal-k\sal) \int d^2z\,J_{(\alpha)}\Jbar_{(\alpha)}
\eqn \dels
$$
 The above equation shows that appropriate \ott\ transformations
of a $SU(2)$-WZNW model generate a continuous line of conformal
field theories which are related by marginal perturbations. The
precise relationship is as follows. The \ott\ transformation
involves two of the target space coordinates,  $\th$ and
$\thwt$, on which the background fields do not depend. Global
shifts in these coordinates are therefore commuting isometries
of the backgrounds, giving rise to conserved isometry currents.
If this symmetry is extendable to local shifts with only
holomorphic or anti-holomorphic dependences on the world-sheet
coordinates, as in \chalinv,  then the theory contains a pair
of chiral currents, and
hence,  a marginal operator. Equation \dels\ shows that
perturbations by this operator are reproduced by appropriate
\ott\ transformations involving $\th$ and $\thwt$. The
integrability of the marginal perturbation (which was proved\RCHOUSCH\ to
all orders in $\alpha$ by working at
the $SU(2)$ point where the conformal structure of the
theory is known)  insures that the new theory
$S_{(\alpha+\delta\alpha)}$ also contains chiral currents and
can be perturbed further. We have proved that an infinite
series of such perturbations can effectively be added up by a
finite \ott\ transformation. This transformation also gives the
expression for the new dilaton field without any extra
calculations.

Next, we consider the $SU(2)\otimes SU(2)$ model defined by the action
$$
S=S_1(\,\phi ,\th ,\thwt\,)+S_2(\,r,t,\twt\,)
\eqn\s
$$
Here $S_1$ is the action given in equation \szero\ with $k=k_1$ and
$S_2$ is obtained from $S_1$ by the replacements $(k_1,\phi,\th,\thwt\,)
\rightarrow (k_2,r,t,\twt\,)$. The backgrounds in $S$ are independent of
the four coordinates $\th,\thwt,t\,\, {\rm and}\,\, \twt$ and can be
transformed
by an $O(4,4)$ group of transformations.
As stated before, a general $O(4,4)$ transformation may be represented by
an $O(4)$ transformation followed by a gauge and general coordinate
transformation.
To study the connection between
these transformations and marginal deformations in the $SU(2)\otimes SU(2)$
model, it is sufficient to consider the action of various $O(2)$ subgroups
of $O(4)$ individually. Note that $O(4)$ contains a subgroup
$O(2)\otimes O(2)$ which transforms each one of the $SU(2)$ theories
separately. The action of this subgroup (followed by an appropriate
scaling of the coordinates) is a trivial extension of the
\ott\ transformation of the $SU(2)$ model considered above and the
$\sigma$-model action of the transformed theory is given by,
$$
S_{(\alpha,\beta)}=S_{1\,(\alpha)}(\,\phi ,\th,\thwt\,)+
S_{2\,(\beta)}(\,r,t,\twt\,)
\eqn\sab
$$
As an example of  the remaining four \ott\ subgroups, we consider the one
which twists  the $\th $ and $t$ coordinates and parametrize the elements
$\Omega_1$, given in \rspara\ , by choosing
$$
R=S^{-1}=\pmatrix{ \cg & 0 & \sg & 0   \cr
                   0   & 1 & 0 & 0     \cr
                   -\sg & 0 & \cg & 0  \cr
                   0 & 0 & 0 & 1   }
\eqn\efawadseven
$$
The transformed backgrounds are obtained from equations \M\ and \mphi\ .
The new metric has conical singularities  at $\phi =0$ and $r=0$ which are
removed by scalings $\th \rightarrow \th /\cg$ and $t \rightarrow t/\cg$,
restoring the periodicities of $\th , \thwt ,t \,{\rm and}\,\twt $ to
$2\pi$. After shifting the $B$-field by a constant, $B_{\th t}\rightarrow
B_{\th t}-\sg \cg$, the transformed $\sigma$-model action and the dilaton
field are given by
$$
\eqalign{
S_{(\gamma)}&=
{1 \over 2 \pi}\int d^2 z \Bigl[ \,{k_1\over 4} \delb \phi \del \phi +
{k_2\over 4} \delb r \del r
\cr & \qquad
+ \Delinv  \Bigl( k_1 \cgsq \sp \delb \th \del \th
+k_2 \cgsq \st \delb t \del t
\cr & \qquad
+k_1(\cgsq+k_1 k_2 \sgsq \st) \cp \delb \thwt \del \thwt
\cr & \qquad
+k_2(\cgsq+ k_1 k_2 \sgsq \sp )\ct \delb \twt \del \twt
\cr  & \qquad
+k_1 \cgsq \cp (\delb \th \del \thwt - \delb \thwt \del \th)
+k_2 \cgsq \ct  (\delb t \del \twt - \delb \twt \del t)
\cr  & \qquad
-k_1 k_2 \sg \cg \sp \st (\delb \th \del t - \delb t \del \th)
\cr & \qquad
-k_1 k_2 \sg \cg \sp \ct (\delb \th \del \twt + \delb \twt \del \th)
\cr  & \qquad
+k_1 k_2 \sg \cg \cp \st (\delb t \del \thwt + \delb \thwt \del t)
\cr & \qquad
+k_1 k_2 \sg \cg \cp \ct (\delb \thwt \del \twt - \delb \twt \del \thwt)
\Bigr) \Bigr]    \cr
\Phi&=-\ln \Delta   \cr   }
\eqn\sgamma
$$
where,
$$
\Delta= \cgsq + k_1 k_2 \sgsq \sp \st
\eqn\efawadeight
$$
For an infinitesimal \ott\ transformation, the change in the action \s\
is
$$
\delta S ={-\gamma \over 2\pi}\int d^2z \left(\,J_1  \Jbar_2-
J_2 \Jbar_1\, \right)
\eqn\efawadnine
$$
where, $J_1, \Jbar_1$ and $J_2, \Jbar_2$ are the chiral $U(1)$ currents of
the
two $SU(2)$ theories in \s. This, clearly, is an integrable
marginal perturbation of the untransformed action $S$. To identify the
theory given by the action $S_{(\gamma)}$ for finite $\gamma$,
we note that the action \sgamma\ has the following chiral
invariances:
$$
\eqalign{
\delta \th &={\cgsq +k_1 k_2 \sgsq \over 2 \cgsq}v(z)\,\,,
\delta \thwt =\half v(z)\,\,,
\delta t=0\,\,, \delta \twt = \half k_1\tan\gamma\, v(z)
\cr
\bar \delta \th &=-{\cgsq +k_1 k_2 \sgsq \over 2\cgsq} \bar v(\bar z)
\,\,,\bar \delta \thwt =\half \bar v(\bar z)\,\,,
\bar \delta t=0\,\,, \bar \delta \twt =-\half k_1 \tan\gamma\,
\bar v(\bar z)
\cr
\delta \th &=0\,\,,  \delta \thwt =-\half k_2 \tan\gamma\,  u(z)\,\,,
\delta t={\cgsq +k_1 k_2 \sgsq \over 2\cgsq} u(z)\,\,,
\delta \twt = \half  u(z)
\cr
\bar \delta \th &=0\,\,,\bar \delta \thwt =\half k_2 \tan \gamma\,
\bar u(\bar z)\,\,,
\bar \delta t=-{\cgsq +k_1 k_2 \sgsq \over 2\cgsq} \bar u(\bar z)
\,\,, \bar \delta \twt = \half \bar u(\bar z)
\cr }
\eqn\chgamma
$$
These invariances give rise to the following conserved chiral currents
$$
\eqalign{
J_{1\,(\gamma)}&=\Delinv\bigl[(\cgsq +k_1 k_2 \sgsq \st)\,J_1
+\sg \cg k_1 \cp  J_2\, \bigr]
\cr
\Jbar_{1\,(\gamma)}&=\Delinv\bigl[(\cgsq +k_1 k_2 \sgsq \st)\,\Jbar_1
-\sg \cg k_1 \cp \Jbar_2\, \bigr]
\cr
J_{2\,(\gamma)}&=\Delinv\bigl[(\cgsq +k_1 k_2 \sgsq \sp)\,J_2
-\sg \cg k_2 \ct  J_1\, \bigr]
\cr
\Jbar_{2\,(\gamma)}&=\Delinv\bigl[(\cgsq +k_1 k_2 \sgsq \sp)\,\Jbar_2
+\sg \cg k_2 \ct \Jbar_1\, \bigr]
\cr  }
\eqn\currgamma
$$
In terms of these currents, the variation of $S_{(\gamma)}$ under a
small \ott\ transformation is given by
$$
\eqalign{
S_{(\gamma+\delta \gamma)}&=S_{(\gamma)}+
{\delta \gamma \over 2\pi (\cgsq+k_1 k_2 \sgsq)^2}
\int d^2 z \bigl[
2\sg \cg(k_2 J_{1\,(\gamma)} \Jbar_{1\,(\gamma)}
         + k_1 J_{2\,(\gamma)} \Jbar_{2\,(\gamma)})
\cr & \qquad \qquad
-(\cgsq-k_1 k_2 \sgsq)(J_{1\,(\gamma)}\Jbar_{2\,(\gamma)}
      -J_{2\,(\gamma)} \Jbar_{1\,(\gamma)})
\bigr]  \cr }
\eqn\delsgamma
$$
This shows that in the continuous one-parameter family of conformal
field theories described by the action $S_{(\gamma)}$,
two adjacent theories are related by a marginal perturbation constructed
from the product of a holomorphic and an anti-holomorphic current.

The action \sgamma\ was obtained by an \ott\ transformation which twists
the coordinates $\th$ and $t$. The full transformation group also contains
three other subgroups which mix the two $SU(2)$ theories in \s\ . The
corresponding transformed actions can be directly obtained from \sgamma\
by the following replacements:
$$
\eqalign{
\th - \twt \,\,\,{\rm twisting}\,\,&:\,\,\, r\rightarrow r+\pi,\,
t\leftrightarrow\twt      \cr
\thwt - t \,\,\,{\rm twisting}\,\, &:\,\,\,\phi \rightarrow \phi+\pi,\,
\th \leftrightarrow \thwt              \cr }
$$
By combining the above two replacements, one obtains the transformed action
for $\thwt-\twt$ twisting.

The result can be generalised for WZNW models based on other groups $G$.
If $r$ is the rank of the group, and $N$ denotes the total number of
generators, then we can choose a parametrisation of the group element (at
least locally) of the form:
$$
g = exp(i\sum_{i=1}^r \theta_{iL}H_i) exp(i\sum_{a=1}^{N-2r}\alpha_a T_a)
exp(i\sum_{j=1}^r\theta_{jR}H_j)
\eqn\egroupelement
$$
where $H_j$ are the generators of the Cartan subalgebra, and $\{T_a\}$
denote a specific set of $N-2r$ generators outside the Cartan subalgebra.
In this case, when the action of the WZNW theory is written in the form of
the conventional $\sigma$-model action in two dimensions, the background
metric and the antisymmetric tensor field components will be independent
of the $2r$ coordinates $\theta_{iL}$, $\theta_{iR}$ ($1\le i\le r$).
Thus there is an $O(2r,2r)$ transformation which can generate new
conformally invariant background, and we would expect that all the $r^2$
marginal deformations generated by taking products of holomorphic and
anti-holomorphic currents in the Cartan subalgebra will be generated by
the $O(2r,2r)$ transformation.

\chapter{$O(3,3)$ Transformation of $SU(2)\otimes (SU(2)/U(1))$ Model}

\def\tt{\tilde t}
\def\ta{\tilde\alpha}
\def\tb{\tilde\beta}
\def\tg{\tilde\gamma}

In this section we shall start with a conformal field theory that is a
product of an $SU(2)$ WZNW model and an $SU(2)/U(1)$ coset model; and then
construct the most general model obtained by $O(3,3)$ transformation of
this model.
The starting model has the metric\RWIT:
$$
ds^2 = k_2 ({1\over 4}dr^2 +\sin^2{r\over 2} dt^2 +\cos^2{r\over 2} d\tt^2 )
+k_1 ({1\over 4}d\phi^2 +\tan^2{\phi\over 2} d\theta^2)
\eqn\ethreeone
$$
and the dilaton and the antisymmetric tensor fields:
$$
\Phi = -\ln\cos^2{\phi\over 2}, ~~~~ B_{t\tilde t} =-k_2\sin^2{r\over 2}
\eqn\ethreetwo
$$
with all other components of the anti-symmetric tensor field being zero.
The fields are independent of the three coordinates $\phi$, $t$ and $\tt$,
and hence we can generate other conformally invariant background via an
$O(3,3)$ transformation in general.
Let us define,
$$
x^1 =\theta, ~~~~ x^2 = t, ~~~~ x^3=\tt
\eqn\ethreethree
$$
As discussed in sect.2, new conformally invariant background can be
generated from the one given in eqs.\ethreeone, \ethreetwo\ via $O(3,3)$
transformation.
A general $O(3,3)$ transformation can be written
as a product of an
$O(3)$ transformation and a three dimensional general
coordinate transformation and gauge transformation involving the
antisymmetric tensor field.
We shall first perform the $O(3)$ transformation of the
background given in eqs.\ethreeone, \ethreetwo.
This is generated by matrices of the form \rspara\ with $R=S^T$.
In the present case ($d=3$), a general $R$ may be taken to be of the form:
$$
R = R_3 R_2 R_1
\eqn\ethreeeight
$$
where,
$$
R_1 = \pmatrix{\cos\ta & -\sin \ta & 0\cr \sin\ta & \cos\ta & 0\cr 0 & 0 &
1\cr}
{}~~ R_2 =\pmatrix{\cos\tb & 0 & \sin\tb\cr 0 & 1 & 0\cr -\sin\tb & 0 &
\cos\tb\cr}
{}~~R_3 = \pmatrix{ 1&0&0\cr 0&\cos\tg&-\sin\tg\cr 0&\sin\tg&\cos\tg\cr}
\eqn\ethreenine
$$
Calculation of the transformed fields is relatively straightforward.
It turns out that the final metric obtained this way is singular at $r=0$,
$r=\pi$ and $\phi=0$ if we assume conventional periodicities in the
variables $\theta$, $t$ and $\tt$.
In order to remove these singularities, we need to perform a linear set of
coordinate transformations of the form:
$$
\pmatrix{\theta\cr t\cr\tt\cr} = A \pmatrix{\theta'\cr t'\cr\tt'\cr}
\eqn\ethreeten
$$
where $A$ is a $3\times 3$ matrix, such that the transformed metric
satisfies the following conditions near $r=0$, $r=\pi$ and $\phi=0$:
$$\eqalign{
{\rm For}~r\simeq 0~~~ & G'_{tt}\simeq k_2 {r^2\over 4}\cr
& G'_{t\tt}, G'_{t\theta} \propto r^2\cr
{\rm For}~r\simeq \pi~~~ & G'_{\tt\tt}\simeq k_2{(r-\pi)^2\over 4}\cr
& G'_{\tt t}, G'_{\tt\theta}\propto (r-\pi)^2\cr
{\rm For}~\phi\simeq 0~~~ & G'_{\phi\phi}\simeq k_1{\phi^2 \over 4}\cr
&G'_{\phi t}, G'_{\phi\tt} \propto {\phi^2}\cr
}
\eqn\ethreeeleven
$$
Such a metric is non-singular if we assume $\theta'$, $t'$ and $\tt'$ to
be angular coordinates with period $2\pi$ each.

It turns out that these requirements fix the matrix $A$ completely.
The dilaton,  metric and the antisymmetric tensor field
components after the transformation are
given by (for convenience of writing we have dropped the primes),
$$
\eqalign{
\Phi = & -\ln\Delta\cr
G_{\theta\theta} &=\Delta^{-1} k_1\sin^2{\phi\over 2}
\{ A^2 \cos^2{r\over 2}
+ \sin^2{r\over 2}\}
\cr
G_{tt} &= \Delta^{-1} k_2\sin^2{r\over 2} \{ B^2
\sin^2{\phi\over 2} + A^2 \cos^2{\phi\over 2}\}
\cr
G_{\tt\tt} &=\Delta^{-1} k_2\cos^2{r\over 2} \{ C^2 \sin^2{\phi\over 2}
+ \cos^2{\phi\over 2}\}
\cr
G_{\theta t} &= \Delta^{-1} \sqrt{k_2 k_1} B \sin^2{\phi\over
2}\sin^2{r\over 2}
\cr
G_{\theta \tt} &= \Delta^{-1} \sqrt{k_2 k_1}
 AC \sin^2{\phi\over 2}\cos^2{r\over 2}
\cr
G_{t\tt} &=0
\cr
B_{\theta t}&= \Delta^{-1} \sqrt{k_2 k_1} AC
\sin^2{\phi\over 2}\sin^2{r\over 2}
\cr
B_{\theta\tt} &= \Delta^{-1} \sqrt{k_2 k_1} B \sin^2{\phi\over
2}\cos^2{r\over 2}
\cr
B_{t\tt} &=- \Delta^{-1} k_2\sin^2{r\over 2} \{\cos^2{\phi\over 2}
+C^2 \sin^2{\phi\over 2}\}
\cr }
\eqn\backabg
$$
where,
$$\eqalign{
A =& {\cos\ta\cos\tb\cos\tg\over k_2\sin\tg -\cos\ta\cos\tb\cos\tg}\cr
B=&\sqrt{k_2 k_1} {\sin\tb\cos\tg \over k_2\sin\tg
-\cos\ta\cos\tb\cos\tg}\cr
C =&\sqrt{k_2 k_1} {\sin\ta\cos\tb\cos\tg\over k_2\sin\tg
-\cos\ta\cos\tb\cos\tg}\cr
\Delta =& \cos^2{\phi\over 2}\{
\sin^2{r\over 2} + A^2 \cos^2{r\over 2}\}
+ \sin^2{\phi\over 2}\{ B^2 \cos^2{r\over 2}
+ C^2 \sin^2{r\over 2}\}
\cr
}
\eqn\eabcdelta
$$
Various special cases of this solution have been discussed in
refs.\RHORAVA\RVENBH.
In the next section we shall see that these models represent conformal
field theories obtained after gauging a marginally deformed $SU(2)\otimes
SU(2)$ WZNW theory.

\def\odd{$O(d,d)$}
\def\ott{$O(2,2)$}
\def\thl{\theta_L}
\def\thr{\theta_R}
\def\th{\theta}
\def\thwt{\widetilde\theta}
\def\tht{\tilde\theta}
\def\twt{\widetilde t}
\def\tt{\tilde t}
\def\cp{\cos^2 {\phi\over 2}}
\def\ct{\cos^2 {r\over 2}}
\def\sp{\sin^2 {\phi\over 2}}
\def\st{\sin^2 {r\over 2}}
\def\cosal{\cos \alpha}
\def\sal{\sin \alpha}
\def\cosbe{\cos \beta}

\def\cg{\cos \gamma}
\def\sg{\sin \gamma}
\def\cgsq{\cos ^2 \gamma}
\def\sgsq{\sin ^2 \gamma}
\def\Delinv{{1\over \Delta}}
\def\d{D^{-1}}
\def\del{\partial}
\def\delb{\bar \partial}
\def\Jbar{{\bar J}}
\def\wij{\del_{[i}w_{j]}}
\def\e{\epsilon}
\def\half{{1\over 2}}
\def\g{{\scriptstyle\rm gauged}}
\def\alt{\alpha}
\def\blt{\beta}
\def\clt{\gamma}
\chapter{Gauging the \odd\ Transformed WZNW-Models:}

In this section we describe the gauging of commuting isometries of a
non-linear $\sigma$-model with chiral symmetries and in which the
background metric, antisymmetric tensor and dilaton fields
do not depend on $d$ of the coordinates. For isometries which involve
these $d$ coordinates, we show that, upto a total derivative term, the
gauged action can be obtained by a covariant derivative replacement. We
then
apply the procedure to gauge the marginally deformed $SU(2)$ and
$SU(2)\otimes SU(2)$ models of section 2 and comment on the results.
This  gives us a method to construct classes of solutions of the
string theory low-energy equations of motion as exact conformal field
theories.

Consider a non-linear $\sigma$-model with a Wess-Zumino term
given by the action
$$
S={1\over 2\pi}\int d^2 z (\,g_{ij}+b_{ij}\,)\delb X^i\del X^j
\eqn\nls
$$
The model may also contain a background dilaton field, $\Phi$, which does
not explicitly appear in the action. The above action is
invariant under a global transformation
$$
\delta X^i =v^\alt \xi^i_\alt(X)
\eqn\iso
$$
if $\nabla_i \xi_{j,\alt}+\nabla_j \xi_{i,\alt}=0\,,\xi^i_\alt\del_i
\Phi=0 $\ and provided there exists a 1-form $K_\alt=K_{i,\alt}dX^i$
given by
$$
b_{ik}\,\del_j\xi^k_\alt+b_{kj}\,\del_i\xi^k_\alt+ \del_k b_{ij}\,
\xi^k_\alt =\del_i K_{j,\alt}-\del_j K_{i,\alt}
\eqn\defk
$$
The isometry currents associated with the above invariance are
$$
\eqalign{
I_\alt&=[\,(\,g_{ij}+b_{ij}\,)\xi^i_\alt-K_{j,\alt}\,]\del X^j \cr
\bar I_\alt&=[\,(\,g_{ij}-b_{ij}\,)\xi^i_\alt+K_{j,\alt}\,]\del X^j \cr}
\eqn\isocur
$$
with $\delb I_\alt+\del \bar I_\alt=0$. In refs. \RHS\RJJM\ it was shown
that  to gauge the isometry \iso, the following conditions must
be satisfied
$$
\eqalign{
b_{jk}\,\xi^k_\blt\del_i\xi^j_\alt+\del_j(b_{ik}\,\xi^k_\blt)\xi^j_\alt
&+\del_j K_{i,\blt}\xi^j_\alt+ K_{j,\blt}\del_i\xi^j_\alt \cr
=&f^\clt_{\alt \blt}(\,b_{ij}\xi^j_\clt+K_{i,\clt}) \cr
K_{i,\alt }\xi^i_\blt=&-K_{i,\blt}\xi^i_\alt   \cr}
\eqn\cond
$$
We restrict ourselves to commuting isometries and set $f^\clt_{\alt \blt}
=0$. If the gauge field $A^\alt _\mu$ transforms as $A^\alt _\mu
\rightarrow A^\alt _\mu+\del_\mu v^\alt $, then the gauge invariant
action is given  by
$$
\eqalign{
S^{\g}&={1\over 2\pi}\int d^2 z[\,(\,g_{ij}+b_{ij}\,)
\delb X^i\del X^j -A^\alt \bar I_\alt -\bar A^\alt  I_\alt  \cr &\qquad
+((\,g_{ij}+b_{ij}\,)\xi^i_\alt \xi^j_\blt-\half(K_{i,\alt }\xi^i_\blt
-K_{i,\blt}\xi^i_\alt ))\bar A^\alt  A^\blt\,] }
\eqn\sgauged
$$
This action has an arbitrariness stemming from the fact that
equations \defk\ and \cond\ do not determine $K_{i,\alt }$ completely.
Moreover, it cannot in general be obtained from \nls\ by a
covariant derivative replacement.

To fix the arbitrariness of the gauged action, we note that for general
$K_{i,\alt}$, the action \sgauged\ does not necessarily describe a
conformal field theory even if the field theory corresponding to the
ungauged action \nls\ is conformally invariant.
However, if the action \nls\ has chiral symmetries, and if the gauge
fields in \sgauged\ couple to the corresponding chiral currents, then the
resulting theory may be described as a coset of the original CFT by $U(1)$
current algebra theories, and hence describes a new conformal field
theory.
The \odd\
transformed WZNW models considered in sect.2 still have residual chiral
invariance given in
eqs.\chalinv, \chgamma.
This enables us to break up the transformation \iso\ into
$$
\delta_L X^i=v^a_L \xi^i_{La}\,,\qquad \delta_R X^i=v^a_R
\xi^i_{Ra}
\eqn\chiso
$$
such that its chiral extension with $v^a_L(z)$ and
$v^a_R(\bar z)$ is also a symmetry of the ungauged action. The
isometry currents $I^\mu_{La}$ and $I^\mu_{Ra}$ associated
with \chiso\ are obtained from \isocur\ on replacing the pair
$(\xi^i_\alt,K_{i,\alt})$ by $(\xi^i_{La},L_{i,a})$ and
$(\xi^i_{Ra},R_{i,a})$  respectively. If we choose to gauge a
diagonal subgroup of \chiso\ with $v^a_L=v^a_R=v^a(z,\bar z)$, and
therefore $A_\mu^{La}=A_\mu^{Ra}\equiv A_\mu^a$,
then the gauged action is given by \sgauged\ with the index $\alpha$
replaced by the index $a$, where, now
$$
\xi^i_a=\xi^i_{La}+\xi^i_{Ra}\,,\qquad K_{i,a}=L_{i,a}+R_{i,a}
\eqn\xik
$$
and, therefore, $I^\mu_a=I^\mu_{La}+I^\mu_{Ra}$.
It can be easily seen that the chiral
invariance of the ungauged action allows us to choose
$$
L_{j,a}=-(\,g_{ij}-b_{ij}\,)\xi^i_{La}\,\qquad
R_{j,a}=(\,g_{ij}+b_{ij}\,)\xi^i_{Ra}
\eqn\lr
$$
satisfying eq.\defk.
This gives $\bar I_{La}=I_{Ra}=0$ and $\del\bar I_{Ra}=\delb I_{La}=0$.
(We shall denote these chiral currents by $J_a$, $\Jbar_a$.)
If this choice is made in the gauged action, the gauge fields
will couple to chiral currents of the ungauged theory, leading to a new
CFT as argued above.
This can be
checked explicitly in the specific examples we have, by verifying that the
backgrounds obtained after
gauge fixing and integrating out the gauge fields satisfy the
$\beta$-function vanishing equations.
This observation has been made earlier in ref.\RROCVER.

Note that after obtaining $L_{j,a}$ and $R_{j,a}$ using eq.\lr, we need to
verify that they satisfy eqs.\cond.
This further restricts the choice of $\xi^i_{aL}$ and $\xi^i_{aR}$ to
anomaly free subgroups.
In all the cases we shall discuss, we shall make appropriate choices of
$\xi^i_{aL}$ and $\xi^i_{aR}$ so that these conditions are satisfied.

Next, we explore the possibility of writing the gauged action \sgauged\ in
terms of
covariant derivatives.
This can be done, provided, by adding appropriate total derivative terms
to the Lagrangian density, we can ensure that the Noether currents
associated with the chiral symmetries are the same as the chiral currents
$J_{a}$, $\bar J_{a}$.
If in \nls\ $b_{ij}$ is
replaced by $b_{ij}+2\wij$ and $\del_\mu X^i$ by $D_\mu X^i=
\del_\mu X^i-\xi^i_a A^a_\mu$, then we get
$$
S^{\g}={1\over 2\pi}\int d^2 z(\,g_{ij}+b_{ij}+2\wij \,)
\bar DX^iDX^j
\eqn\covsg
$$
This is the correct gauged action \sgauged\ provided one can find a
$w_j$ such that
$$
2\wij\, \xi^j_a=L_{i,a}+R_{i,a}
\eqn\w
$$
This is possible if we restrict ourselves to backgrounds of the
form \bac\ and to shift isometries in the coordinates on which
the backgrounds do not depend $(\xi^\alpha_a=0,\,\del_i\xi^m_a=0)$.
Equations \defk\ and \cond\ then imply that $L_{m,a}$ and $R_{m,a}$
are constant while $L_{\alpha,a}=R_{\alpha,a}=0$. $2\wij$ is,
therefore, a constant matrix with non-zero elements only in the
subspace spaned by the coordinates the isometries in which are
gauged. Combining \w\ with
eq. \lr\ gives
$$
2\wij\,\xi^j_a=g_{ij}(\xi^j_{Ra}-\xi^j_{La})-b_{ij}(\xi^j_{La}
+\xi^j_{Ra})
\eqn\wconf
$$

In the following we use equations \covsg\ and \wconf\ to gauge the
shift isometries of \odd\ transformed WZNW models.
First, we
consider the axial gauging of the deformed $SU(2)$ model \salpha.
Reading out $\xi^i_L$ and $\xi^i_R$ from \chalinv\ and
substituting in \wconf, we get $\del_{[\th}w_{\tilde\th]}=0$,
\foot{This is a consequence of the appropriate choice of $L$ in
\l. For vector gauging the proper choice is $L=\cosal (k\cosal +
\sal)-k$. Since the added term is independent of $\alpha$, it
does not affect equation \deltas.} leading to the gauged action
$$
S^{\g}_{(\alpha)}(\phi,\th,\thwt,A)=S_{(\alpha)}(\phi,\th,\thwt)
-{1\over 2\pi}\int d^2 z\bigl(\,A\Jbar_{(\alpha)}+\bar A
J_{(\alpha)}-{k\cp\over \Delta\cos^2\alpha}A \bar A \,\bigr)
\eqn\salphag
$$
where $J_{(\alpha)}$, $\bar J_{(\alpha)}$ have been defined in eq.\curral.
Surprisingly, after gauge fixing ($\thwt=0$) and integrating out the gauge
field, one obtains the $SU(2)/U(1)$ blackhole of the
untransformed theory
$$
\eqalign{
ds^2&=(k/4)d\phi^2+k\tan^2(\phi/2)\,d\th^2 \cr
e^{-\Phi}&=\cp\cr}
\eqn\efawadten
$$
The reason for the disappearence of $\alpha$ lies in the freedom
to redefine the gauge fields. In fact, substituting
$$
A=\cosal \sqrt{\Delta} A^\prime + \cosal (\cosal - \sqrt{\Delta})
{J \over k \cp}
\eqn\efawadeleven
$$
in \salphag, with a similar expression for $\bar A$, we find that
$$
S^{\g}_{(\alpha)}(\phi,\th,\thwt,A)=S^{\g}_{(\alpha=0)}
(\phi,\th,\thwt,A^\prime)
\eqn\ss
$$
The Jacobian of the transformation modifies the path integral
measure and readjusts the dilaton field to zero.

This result may also be understood in a qualitative manner by noting that
gauging the deformed
WZNW model corresponds to taking the coset of the conformal field theory
before gauging by the $U(1)$ current algebra theory.
The action of the operators $J_{(\alpha)}$, $\bar J_{(\alpha)}$ becomes
trivial in the coset
theory, and hence one would expect that the effect of perturbing the
original theory by the $J_{(\alpha)}\bar J_{(\alpha)}$ operator before
gauging the theory will be washed out after gauging.
This, in turn, implies that the final theory will be independent of the
parameter $\alpha$, which is indeed the case here.

Before proceeding further, we use equation \ss\ to clarify the meaning of
the observations made in refs.\RALOK\ and \RVENBH. In \RALOK\ it was
observed that the gauged action of ref.\RGIVROC\
can be obtained from the corresponding ungauged action by a
constant \odd\ transformation. In the present context, equation
\salpha\ , before performing the scalings following \efawadfour , shows
that an \ott\ transformation of the $SU(2)$ model with $\alpha=\pi/2$ gives
the $SU(2)/U(1)$ coset model and a free field $\thwt$. To
understand why the coset model is obtained by such an \ott\
transformation, it is sufficient to look for some $\alpha
=\alpha_0$ for which $\thwt$ decouples as a free field
$$
S_{(\alpha=\alpha_0)}(\phi,\th,\thwt)=S^\prime_{(\alpha=\alpha_0)}
(\phi,\th)
+{1\over 2\pi} \int d^2z \delb \thwt \del \thwt
\eqn\efawadtwelve
$$
Now, quotienting by the axial $U(1)$ subgroup simply eliminates
$\thwt$ on the right hand side, leaving
$S^\prime_{(\alpha=\alpha_0)}(\phi,\th)$.
On the other hand, this, by \ss, must be the same as $S^{\g}_{(\alpha=0)}$.
This shows that once we know that $S_{\alpha=\alpha_0}$ can be written in
the form of eq.\efawadtwelve, then $S'_{\alpha=\alpha_0}(\phi,\theta)$
must be the action of the gauged WZNW model ($SU(2)/U(1)$ coset model).
Though the situation considered in ref.\RALOK\ is more general, the
results can be understood in the same way. The above argument is valid
provided the \odd\ transformed theory is written in a form that can be
gauged by covariant derivative replacement. In ref.\RVENBH\ it was
observed that by adding a free field to the singular $SL(2,R)/U(1)$
background and an \ott\ transformation, one can "boost away" the
singularity. This can be  understood by inverting the argument based on
eqn \efawadtwelve\ : the singularities which are boosted away are the ones
that appear as a consequence of a $U(1)$  gauging of the boosted theory,
for which \ss\ holds. The argument can also be generalized to the case
of the 4-dimensional solution of ref.\RWIT\ considered in \RVENBH .

Next, we consider the gauging of the $O(4,4)$ transformed $SU(2)\otimes
SU(2)$ model in
which the two $SU(2)$ sectors have not been mixed by $O(4,4)$.
As shown above, a naive
gauging of a $U(1)$ subgroup belonging to either of the two $SU(2)$'s
will simply give the direct
product of  a marginally deformed $SU(2)$ WZNW model and an $SU(2)/U(1)$
coset model.
To obtain a non-trivial result,
we gauge a $U(1)$ subgroup which acts simultaneously on both
sectors of the theory.
The action of this subgroup is given by
$$
\delta \th =0\,,\quad \delta \thwt ={v(z,\bar z)\over \cos^2\alpha}\,,
\quad \delta t =0\,,\quad \delta \widetilde t =\lambda
{v(z,\bar z)\over \cos^2\beta}
\eqn\efawadthirteen
$$
where, $\lambda$ is a new parameter. The gauged action
is given by
$$
\eqalign{
S^{\g}_{(\alpha ,\beta)}&=S_{(\alpha)}(\,\phi,\th,\thwt\,)
+S_{(\beta)}(\,r,t,\widetilde t\,)
-{1\over 2\pi}\int d^2 z\Bigl[\,A\, (\Jbar_{1\,(\alpha)}+
\lambda\, \Jbar_{2\,(\beta)}) \cr
&\qquad +\bar A\, (J_{1\,(\alpha)}+
\lambda\, J_{2\,(\beta)})
-A\bar A\, (\,{\cp\over\cos^2\alpha \Delta_1}+\lambda^2
{\ct\over\cos^2\beta \Delta_2}\,)\,\Bigr] \cr }
\eqn\sabg
$$
where, the subscripts $1$ and $2$ refer to the theories
$(\,\phi,\th,\thwt\,)$ and $(\,r,t,\widetilde t\,)$, respectively.
Choosing a gauge $\thwt=0$ and integrating out the gauge field,
we obtain the following background fields:
$$
\eqalign{
G_{\th \th}&=\d k_1 \sp (\st+Q \ct)           \cr
G_{tt}&=\d k_2 \st (k_1 k_2 P^2 \sp + Q \cp)  \cr
G_{\tt \tt}&=\d k_2 \cp \ct                   \cr
G_{\th t}&=\d k_1 k_2 P \sp \st               \cr
G_{\th \tt}&=G_{t\tt}=B_{\th t}=0             \cr
B_{\th \tt}&=\d k_1 k_2 P \sp \ct             \cr
B_{t \tt}&=-\d k_2 \st \cp                    \cr
\Phi&=-\ln D                                  \cr  }
\eqn\backab
$$
where,
$$
\eqalign{
D&=\cp + k_1 k_2 P^2 \ct -(1-Q + k_1 k_2 P^2) \cp \ct  \cr
P&={\lambda \over k_1} \left(\cosal \over \cosbe \right)^2  \cr
Q&= (1-k_2\tan\beta)^2+\lambda^2 {k_2\over k_1} \left(\cosal \over
\cosbe \right)^4 (1-k_1 \tan \alpha)^2  \cr }
\eqn\dpq
$$
These fields depend on only two independent parameters $P$ and $Q$. The
reason for the disappearance of the third parameter is again to be sought
in the freedom to redefine the gauge fields. In fact, if $A$ and $\bar A$
are redefined such that the terms linear in the gauge fields in \sabg\ are
eliminated, the resulting action will depend only on  $P$ and  $Q$. Now
we can compare the backgrounds in \backabg\ and \backab . In fact, if we
put $ C =0 $ in \backabg, and identify $P$ and $Q$ of \backab\ as
$$
P = {B\over \sqrt{k_2 k_1}}\,,\,\,\,
Q = A^2
\eqn\pq
$$
then the two sets of background fields in \backabg\ and \backab\ turn out
to be the same.

To generate a class of solutions \backabg\ with non-zero $C$, we cosider
the gauging of the marginally deformed $SU(2)\otimes SU(2)$ theory
obtained by an \ott\ which twists $\th$ and $\twt$.\foot{The background
fields obtained on gauging the $\th-t$  twisted model are still of the
form \backab.} The action $S_{(\gamma)}'$, chiral symmetries, and
chiral currents, $J_{i\,(\gamma)}'$ and $\Jbar^\prime_{i\,(\gamma)}$,
of this theory are obtained from equations \sgamma,\chgamma\ and
\currgamma\ by the replacements $r\rightarrow r+\pi$ and $t\leftrightarrow
\twt$. The transformation to be gauged is $\delta \th =0,\, \delta \thwt
=v(z,\bar z),\, \delta t= \lambda v(z,\bar z)\, {\rm and}~ \delta \twt =0$.
{}From \wconf\ we get $\partial_{[\th}w_{\tht]}=-k_1,\, \p_{[t}
w_{\tt]}=k_2$ which, using \covsg,
leads to the following gauged action
$$
\eqalign{
S'^{\g}_{(\gamma)}&=S_{(\gamma)}'-{1\over 2\pi}\int d^2 z\Bigl[
\,A\, (\Jbar_{1\,(\gamma)}'+\lambda\, \Jbar_{2\,(\gamma)}')
+\bar A\, (J_{1\,(\gamma)}'+ \lambda\, J_{2\,(\gamma)}')
\cr & \qquad \qquad
-\Delinv A\bar A\, \Bigl(\,k_1\cp \,(\cgsq +k_1 k_2 \sgsq \,\ct )
\cr & \qquad \qquad
+\lambda^2 \,k_2 \st \,(\cgsq +k_1 k_2 \sgsq \,\sp)\,\Bigr)
\Bigr]                 \cr }
\eqn\sgammag
$$
where,
$$
\Delta =\cgsq+k_1 k_2 \sgsq \sp \ct
\eqn\delprime
$$
On fixing the gauge $\thwt =0$ and integrating out the gauge fields, we
obtain the following expressions for the metric, antisymmetric tensor
field and the dilaton field \foot{These fields can also be obtained from
\backab\ after replacing $r$ by $r+\pi$ and interchanging $t$ and $\twt$.
This corresponds to gauging a combination of axial and vector $U(1)$'s.}:
$$
\eqalign{
G_{\th \th}&=\d k_1 \sp [(1+k_1 k_2 M N^2)\st+ M \ct)              \cr
G_{tt}&=\d k_2 M \st \cp                                           \cr
G_{\tt \tt}&=\d k_2 \ct [k_1 k_2 M N^2 \sp +(1+k_1 k_2 M N^2)\cp ] \cr
G_{\th \tt}&=\d k_1 k_2 M N \sp \ct     \cr
G_{\th t}&=G_{t\tt}=B_{\th \tt}=0       \cr
B_{\th t}&=\d k_1 k_2 M N \sp \st      \cr
B_{t \tt}&=\d k_2 M \cp \ct             \cr
\Phi&= -\ln D                           \cr }
\eqn\backg
$$
where,
$$
\eqalign{
D&= k_1 k_2 M N^2 +\cp -k_1 k_2 M N^2 \ct +(M-1) \cp \ct     \cr
M&={\cgsq + k_1 k_2 \sgsq \over \cgsq - \lambda^2 k_2 \sgsq}\,,
\qquad  N=\lambda/k_1    \cr   }
\eqn\dmn
$$
To compare the two sets of background fields in \backabg\ and \backg\ (up
to an overall shift of the dilaton field),
we have to set $B=0$ and $B_{t\tt}\rightarrow B_{t\tt}+k_2$ in \backabg\
and identify its remaining parameters in terms of the parameters $M$ and
$N$ of \backg\ as
$$
M= {A^2\over 1-C^2}\, , \,\,\,
N = {C\over A\sqrt{k_2 k_1}}
\eqn\mn
$$
This shows that various classes of solutions of the low energy equations
of motion which are obtained by the action of the $O(3,3)$ group on the
$\big(SU(2)/U(1)\big)\otimes SU(2)$ coset model, can be constructed as
$U(1)$
cosets of the marginally deformed $SU(2)\otimes SU(2)$ WZNW theory.

\chapter{Summary and Conclusion}

In this paper we have studied $\sigma$-models obtained by \odd\
transformations of WZNW models and have shown that they correspond to
finite marginal deformations of the original WZNW models. These marginal
deformations are generated by a product of holomorphic and
anti-holomorphic currents belonging to the Cartan subalgebra of the
underlying current algebra. We have also studied the gauging of $U(1)$
subgroups of the marginally deformed theory and have shown that the
results can be obtained by $O(d,d)$ transformations of the gauged
unperturbed WZNW model.

Our analysis provides a way to give a $\sigma$-model description of
marginally deformed WZNW models and their cosets. The existance of such
deformed models was proved to all orders in perturbation theory
in \RCHOUSCH ,where, it was shown that marginal perturbations by products
of holomorphic and anti-holomorphic currents from the Cartan subalgebra are
integrable. Though, it is easy to write down the form of these models to
first order in the perturbation parameter, obtaining the $\sigma$-model
for a finite deformation is in no way a straightforward task without the
help of the $O(d,d)$ transformations.

Although the form of the $\sigma$-model actions that we have derived for
the deformed WZNW models and their gauging is valid to all orders in the
deformation parameter, the expression is valid only to the lowest order
in the derivatives (in the target space).
In other words, the $\sigma$-model $\beta$-functions are zero only to one
loop order, or in the large $k$ limit.
This limitation stems from the fact that the explicit $O(d,d)$
transformation laws of various fields are known only to this order.
The existence of an $O(d,d)$ transformation that converts a conformally
invariant background to another conformally invariant background has,
however, been established to all orders\RSEN\RFAWAD. This, in turn, shows
the existence of  $\sigma$-models which represents deformed WZNW models
(and their gauging) to all orders in the $\sigma$-model loop expansion.

\refout
\end